\documentclass[manuscript]{acmart}
\AtBeginDocument{%
  }

\usepackage{xcolor}
\usepackage{subcaption}
\newbool{debug}
\booltrue{debug}
\usepackage{xcolor}
\definecolor{rahulColor}{RGB}{0, 0, 139}
\ifbool{debug}{
\newcommand{\reds}[1]{\noindent\textcolor{red}{\small \bf [SR: #1]}}
}{
\newcommand{\reds}[1]{}
}
\usepackage[normalem]{ulem}

\usepackage{tabularx}
\usepackage{booktabs}
\usepackage{makecell}
\usepackage{array}
\setcopyright{none}
\renewcommand\footnotetextcopyrightpermission[1]{}
\begin{document}

\title{KI-Adventskalender: An Informal Learning Intervention for Data \& AI Literacy}


\author{Rahul Sharma}
\email{rahul.sharma@dfki.de}
\affiliation{%
  \institution{German Research Center for AI (DFKI GmbH), }
  \institution{RPTU Kaiserslautern-Landau}
  \city{Kaiserslautern}
  \country{Germany}
}

\author{Lars Henrich}
\affiliation{%
  \institution{RPTU Kaiserslautern-Landau}
  \city{Kaiserslautern}
  \country{Germany}}

\author{Larisa Ivanova}
\affiliation{%
  \institution{German Research Center for AI (DFKI GmbH), }
  \institution{Universität des Saarlandes}
  \city{Kaiserslautern}
  \country{Germany}
}

\author{Arsalan Karimzadmotallebiazar}
\affiliation{%
  \institution{German Research Center for AI (DFKI GmbH)}
  \city{Kaiserslautern}
  \country{Germany}
}

\author{Annette Bieniusa}
\affiliation{%
 \institution{RPTU Kaiserslautern-Landau}
 \city{Kaiserslautern}
 \country{Germany}}
 
\author{Leo van Waveren}
\affiliation{%
  \institution{RPTU Kaiserslautern-Landau}
  \city{Kaiserslautern}
  \country{Germany}
}

\author{Sebastian Vollmer}
\affiliation{%
  \institution{German Research Center for AI (DFKI GmbH), }
  \institution{RPTU Kaiserslautern-Landau}
  \city{Kaiserslautern}
  \country{Germany}}

\renewcommand{\shortauthors}{Sharma et al.}

\begin{abstract}
  Secondary school students increasingly encounter AI systems whose outputs depend on data quality, evaluation choices and modeling assumptions. To provide accessible entry points to these interconnected concepts, we developed
  \textbf{\textit{KI-Adventskalender}}, a free web-based extracurricular initiative 
  with 24 didactically curated, short, guided micro-challenges released daily in December, targeting
  data-centric competencies and socio-technical themes that shape how data are interpreted in practice. 
  Drawing on two annual iterations, we report aggregate platform traces characterizing participation and task-level engagement. Participation increased substantially in 2025, but early attrition persists. Progression stabilized after midpoint: among users reaching Day 12 in 2025, more than 75\% completed the calendar. Competence cluster performance shifted across years; higher revision rates co-occurred with strong pass rates, suggesting sustained engagement.
  We use these observations to motivate a next-step measurement agenda: tighter task instrumentation, embedded micro-assessments and mixed-method evaluation designs that can distinguish persistence from conceptual uptake, knowledge progression and durable learning outcomes. 
\end{abstract}



\begin{teaserfigure}
  \includegraphics[width=\textwidth]{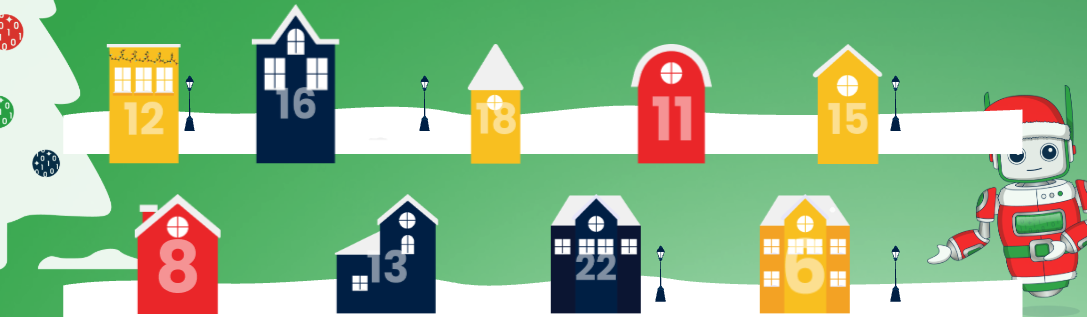}
  \caption{KI-Adventskalender, 2025}
  \Description{A screenshot of the AI Advent Calendar showing colourful doors. An AI challenge hides behind each door.}
  \label{fig:teaser}
\end{teaserfigure}


\maketitle

\section{Introduction}

People encounter AI systems daily---product recommendations, image generators, machine translation, and much more. Using these systems thoughtfully requires data reasoning: evaluating data quality, drawing justified inferences \cite{gould2017data,ridsdale2015strategies} and assessing algorithmic behavior and constraints \cite{10.1145/3313831.3376727,allen2024ed}. 
Data and AI literacy face a common measurement challenge: existing scales vary in what and how they assess \cite{lintner2024systematic,6875906,Brockbank2025} and whether task scores or trace logs indicate understanding  remains contested \cite{10.1145/3544548.3581406,10678815}. The challenge is starker in informal settings where controlled studies are infeasible. 

Secondary education is characterized by an ongoing tension between the need for accessible, motivating entry points and the requirement for conceptual depth that fosters meaningful reasoning. K--12 curricula are intended to foster conceptual knowledge and critical thinking beyond tool use \cite{10.1145/3408877.3432513,YUEYIM2024100319} and specific designs show promise: interactive explainability activities \cite{10.1145/3643834.3660722}, participatory ethics activities \cite{10.1145/3392063.3394396,10.1145/3643834.3661515}, privacy exercises \cite{10.1145/3613904.3642460}. However, most interventions primarily measure engagement, not learning \cite{casal2023ai,LAUPICHLER2022100101,lintner2024systematic} and few evaluate concept progression post-intervention \cite{casal2023ai}. Extracurricular frameworks suggest repeated brief tasks sustain engagement \cite{druga20214as} and reduce cognitive overload \cite{druga20214as}. 
\textit{\textbf{KI-Adventskalender}}\footnote{\scriptsize\href{https://www.ki-adventskalender.de/en}{\texttt{AI Advent Calendar}}} was developed as a low-threshold web-based extracurricular learning platform for the secondary (high school) students, structured as a digital advent calender. Each day from December 1--24 unlocks a new ``door'' containing brief self-contained tasks. Instead of teaching a full curriculum, the 24 doors target key concepts at data/AI intersection: data representation, evaluation metrics, model drift, explainability and fairness. Tasks are intentionally designed to be completed without specialized software or equipment and take 15--20 minutes. We analyze two years (2024--2025) of aggregated platform logs to examine participation, progression and challenge-level engagement.

\textbf{Research questions}: Specifically, we ask: (1) What participation and progression patterns characterize the informal micro-challenge intervention? (2) Do competence clusters exhibit distinct engagement and difficulty signatures? (3) How does structural sequencing of challenges relate to observable performance shifts? (4) What measurement gaps remain for validating data/AI literacy learning?
We do not claim completion or correctness proves learning; we show what behavioral signals look like in this format, where they fail as proxies and what instrumentation would support stronger claims. \textbf{Contributions}: (1) instructional scope of the door set; (2) aggregate patterns from 2024--2025: registration/progression curves, challenge-level difficulty signals, topics with high revision behavior; (3) a measurement agenda for future work: improved event logging, embedded micro-assessments, mixed-methods designs distinguishing persistence from comprehension and durable learning.

\section{Data \& Methods}



\subsection{Platform Mechanics and Interaction Loop}
The platform is a web application accessible via any modern browser on desktop, tablet or mobile. Registration requires email, username, gender, country and referral source; prize-eligible students additionally provide full name, age, school and expected graduation year. Each door unlocks at midnight CET on its scheduled day with a one-week deadline. Answers can be revised unlimited times before the deadline; system scores the last submission after the deadline. Logs capture user ID, door ID, submission timestamp, answer content, and revision count. Time-on-task is calculated from when a user starts the challenge until submission in one session, though users may leave their browser tabs open. 

\subsection{UI/UX and Task Design}
The calendar interface displays 24 numbered boxes in a jumbled up fashion typically seen in advent calendars. Clicking a door loads a challenge page with narrative framing, instructional text, occasionally an embedded diagram or video and an answer input field. Answer formats include single-choice radio buttons (e.g. selecting a fairness metric), multiple-choice checkboxes (e.g., identifying bias sources) and free-text boxes (e.g., entering threshold values or listing algorithm steps). Interactive elements appear in several doors: sliders for adjusting decision thresholds and observing precision-recall shifts, draggable points for \textit{k}-means clustering visualization, before/after image pairs for deepfake detection. No hints are available during the challenge window, though doors occasionally link to external explainer videos or articles. Each door targets one concept without dependencies, requiring no prior AI exposure. Tasks are designed for 15--20 minute completion on any device. 

\subsection{Door Content Scope}
Table~\ref{tab:clusters} summarizes topic distribution. In 2024, doors emphasized foundational ML mechanics (neurons, gradient descent, decision trees, KNN) and introductory framing. In 2025, the design shifted toward data-centric reasoning: data representation and cleaning, binary representations and convolutions, evaluation metrics, precision-recall trade-offs, threshold tuning, model drift, XAI methods, formal fairness definitions; privacy and deepfakes with more technical depth. Both years included genAI (LLMs, GANs) and system-level topics (hardware, agents, collaborative filtering). While the competence clusters remained the same, the order in which the challenges from these clusters are released is different. In year 2024, challenges from a cluster appeared close to each other, while in year 2025 they were interleaved with challenges from other clusters.

\begin{table}[t]
  \centering
  \caption{Competence clusters and door assignments in 2024 and 2025.}
  \label{tab:clusters}
  \footnotesize
  \setlength{\tabcolsep}{4pt}
  \renewcommand{\arraystretch}{1.08}

  \begin{tabularx}{\linewidth}{@{}X
    >{\raggedright\arraybackslash}p{0.18\linewidth}
    >{\raggedright\arraybackslash}p{0.18\linewidth}@{}}
    \toprule
    \textbf{Competence Clusters} & \textbf{2024} & \textbf{2025} \\
    \midrule

    \textbf{Introduction}: basic AI framing, terms, how to read common claims &
    1, 2, 3 &
    1, 10 \\

    \textbf{System level}: model families, pipelines, deployment context, agentic setups &
    5, 7, 8, 10, 11, 12, 14, 15, 24 &
    8, 12, 13, 15, 17, 19, 23, 24 \\

    \textbf{Data representation}: how data are encoded/structured, text \& vision representations &
    9, 16, 20 &
    2, 4, 9, 18 \\

    \textbf{Evaluation}: metrics, thresholds, validation logic, and tuning choices &
    13, 19, 22 &
    6, 7, 20 \\

    \textbf{Reliability}: limitations, error modes, interpretability &
    17, 18 &
    3, 21 \\

    \textbf{Temporal}: time series, drift, distributional change over time &
    23 &
    11, 16 \\

    \textbf{Societal stakes}: fairness, privacy, synthetic media, misuse risks &
    4, 6, 21 &
    5, 14, 22 \\

    \bottomrule
  \end{tabularx}
  \Description{Table showing the seven competence clusters and which challenges were part of such clusters in 2024 and 2025}
\end{table}

\subsection{Data Sources}
We analyze operational traces generated from 2024 and 2025 deployments:

\begin{itemize}
    \item User ID: Anonymized unique identifiers (UUIDs) assigned at registration
    \item Timestamp: Precise submission time for each attempt
    \item Door ID: Challenge identifier (1–24) 
    \item Attempt Count: Number of submissions per user per door
    \item Final Correctness: Binary indicator (1 = Pass, 0/Null = Fail)
\end{itemize}

\subsection{Operational Definitions}
Key metrics are operationalized as follows: 

\begin{itemize}
    \item \textbf{Progression}: defined as the ratio of unique active users on a given day $i$ relative to the number of unique active users on the immediately preceding day $i-1$. It is calculated as $\frac{N_{day_i}}{N_{day_{i-1}}} \times 100$. This metric serves as proxy for longitudinal engagement. 

    \item \textbf{Pass Rate}: defined as the proportion of active participants on a given day who successfully submitted a correct answer. It is calculated as $\frac{P_{correct}}{P_{attempting}}$. This metric proxies the accessibility of solvability of the challenge content. 

    \item \textbf{Mean Attempts}: The average number of submissions required to active a correct result (or before abandonment) for a specific door. A higher mean attempt value is operationalizes as an indicator of higher challenge complexity.

    \item \textbf{Weighted Cluster Means}: Competence cluster-level metrics were calculated using the participation-weighted means. For a cluster $C$ containing a set of days $D$, the weighted mean attempt rate is $\frac{\sum_{d\in D}(MeanAttempts_d \times N_d)}{\sum_{d\in D} N_d}$

    \item \textbf{Unit of Analysis}: The primary unit of analysis is the user-door interaction. Aggregations are performed at the daily level and the cluster level. 
\end{itemize}
\subsection{Analytical Approach}
First, participation and progression are analyzed by plotting cumulative relative active users and progression across the 24-day period to identify attrition points. Second, we assess difficulty via mean attempts and pass rate per door and per competence cluster, visualized using charts and scatter plot matrices to examine relationships between complexity (attempts) and success (pass rate). We characterize quadrants of this matrix to infer user experience states. Third, we compare these metrics across 2024 and 2025 runs.

\subsection{Ethics and Privacy}
Users consent to data collection via terms stating anonymized usage data may inform research. No data are shared with third parties except postal providers for prize delivery. All analyses use aggregate summaries. We do not report individual trajectories, school-level breakdowns or statistics with cell sizes below 5. 



\section{Results}








\subsection{Participation and Progression}
The platform's onboarding efficiency imporved significantly year-over-year. While top-of-funnel registrations were more streamlined, the users who attempted at least one challenge increased by 33.5\% in 2025. 

The participation statistics from both years show a significant expansion of the calendar's initial reach, with the 2025 run initiating with a 47\% higher unique active participants on Day 1 compared to 2024. This growth in early interest translated to a higher overall volume of activity, resulting in an approximate 21\% increase in total attempts across all 24 challenges. Furthermore, the expanded reach of the 2025 iteration translated into higher absolute progression, culminating in a 25.3\% increase in the total number of users who completed the final challenge.

\begin{figure}
  \centering
  \begin{subfigure}[b]{0.49\textwidth}
    \centering
    \includegraphics[width=\textwidth]{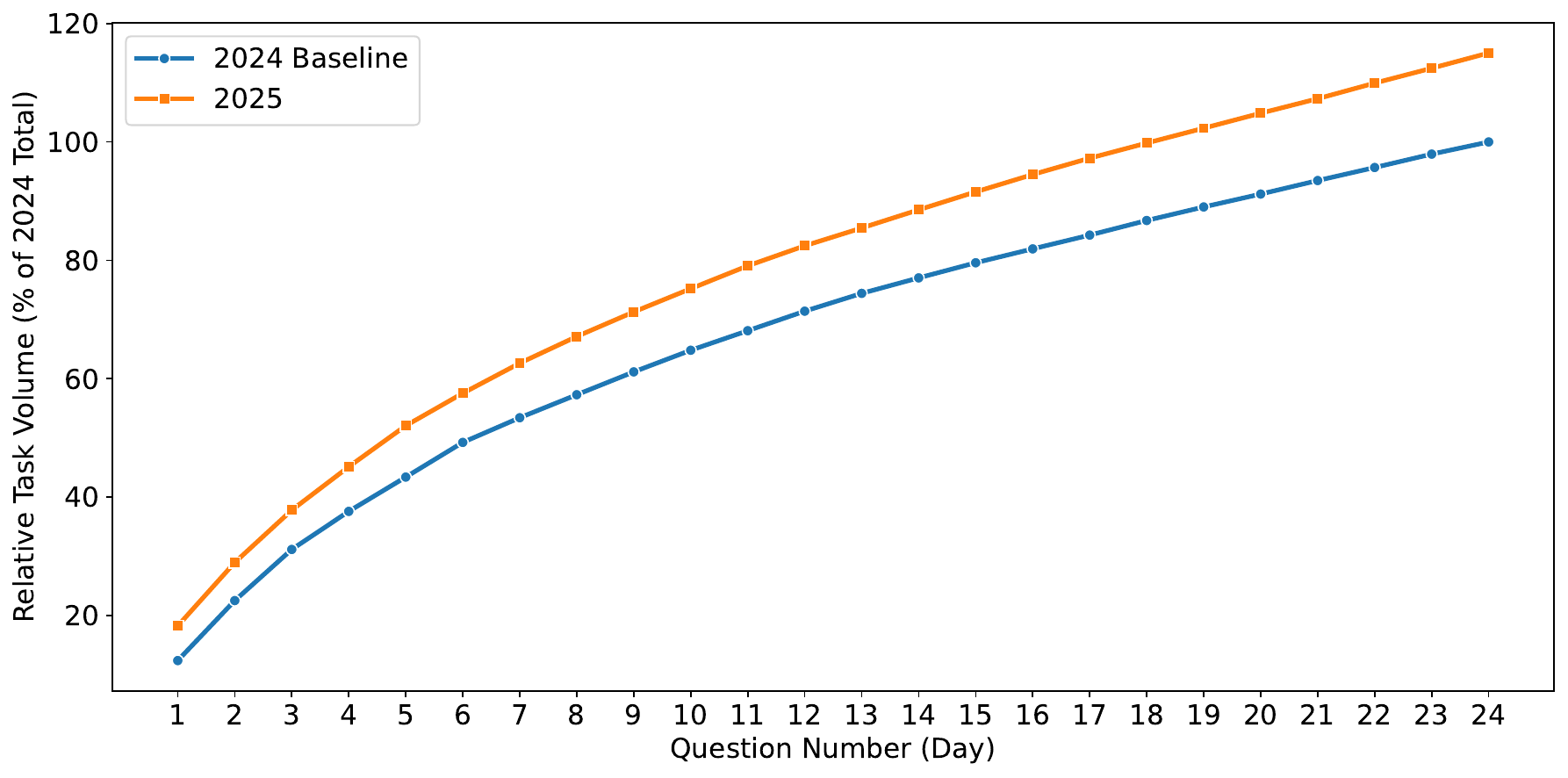}
    \caption{Cumulative relative active users for the years 2024 and 2025}
    \label{fig:cumulative_users}
  \end{subfigure}
  \hfill
  \begin{subfigure}[b]{0.49\textwidth}
    \centering
    \includegraphics[width=\textwidth]{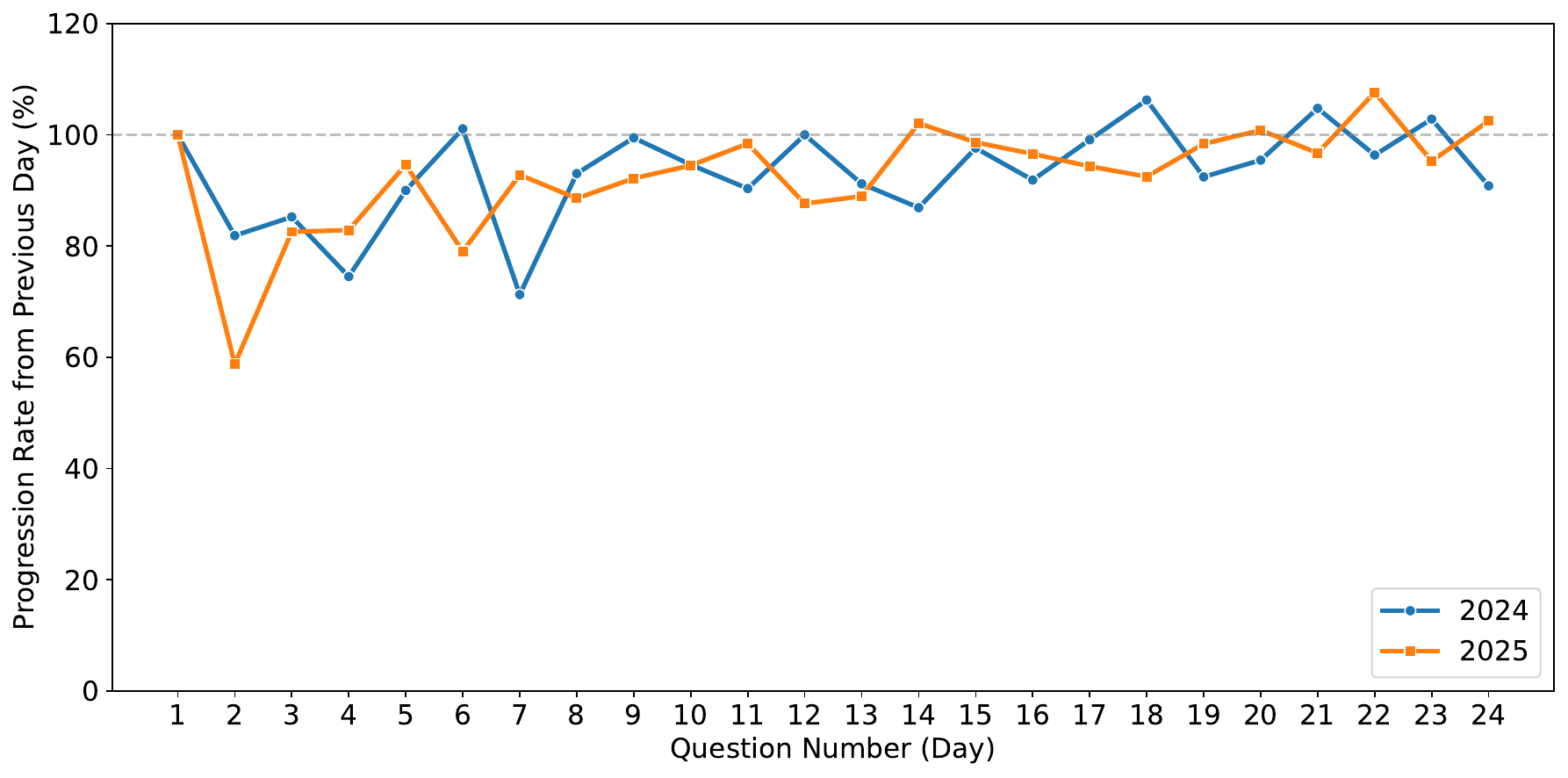}
    \caption{Progression for the years 2024 and 2025}
    \label{fig:conditional_progression}
  \end{subfigure}
  \caption{Cumulative relative active users(a) and progression (b) for both runs of KI-Adventskalender}
  \Description{A two column graph showing the cumulative active users for the years 2024 and 2025 on the left side. The right side graph shows the cumulative progression in both years from challenge 1 to challenge 24.}
  \label{fig:participationAndProgression}
\end{figure}

Regarding longitudinal progression, the two iterations exhibited distinct engagement dynamics. The 2025 run exprerienced a higher initial filtering effect during the first day-to-day transition. However, in both years, once participants reached the midway point (Day 12), they showed high resilience, with progression stabilizing significantly. For those who reached the second half of the calendar, the completion probability was exceptionally high, particularly in 2025, where over 76\% of users active of Day 12 stayed until the final challenge. Figure \ref{fig:participationAndProgression} shows the cumulative relative active users and progression from Day 1 to Day 24 for both runs of KI-Adventskalender.

\subsection{Challenge-level Performance}

\begin{figure}
  \centering
  \begin{subfigure}[b]{0.49\textwidth}
    \centering
    \includegraphics[width=\textwidth]{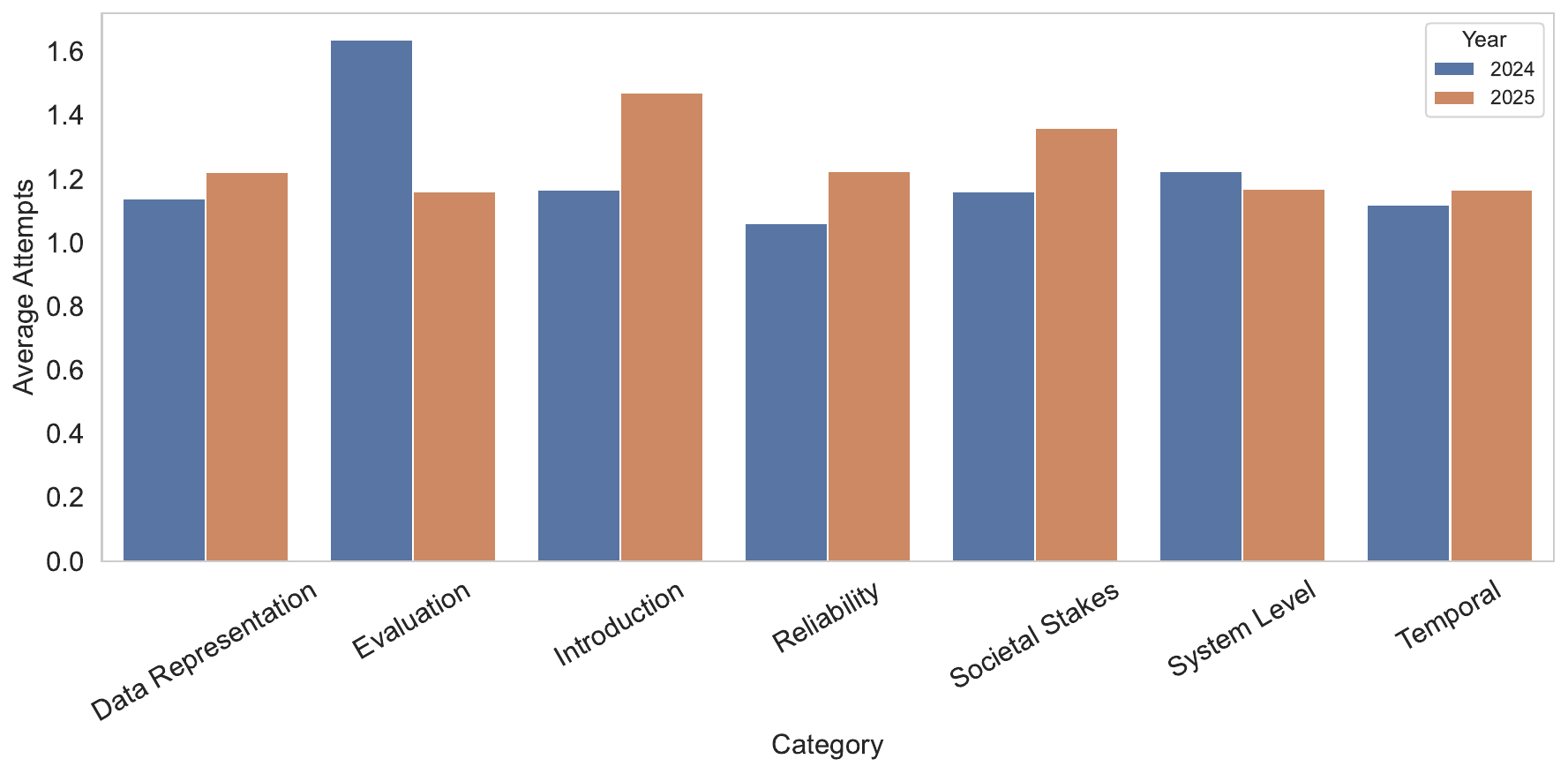}
    \caption{Mean attempts for competence cluster}
    \label{fig:meanAttemptsCategory}
  \end{subfigure}
  \hfill
  \begin{subfigure}[b]{0.49\textwidth}
    \centering
    \includegraphics[width=\textwidth]{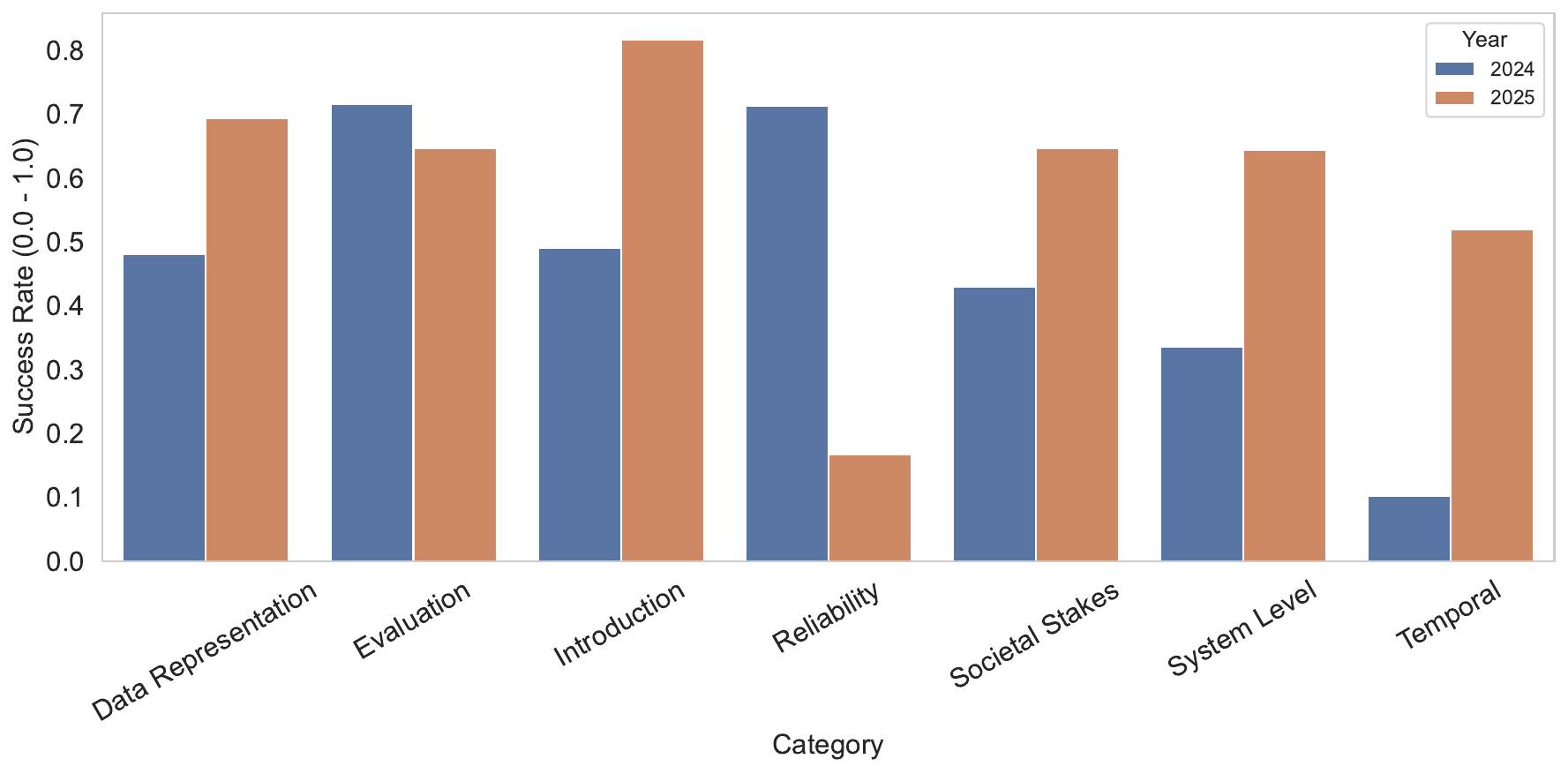}
    \caption{Pass rate for competence cluster}
    \label{fig:passRateCategory}
  \end{subfigure}
  \caption{Mean attempts (a) and rate of passing the challenge  (b) across competence clusters for both runs of KI-Adventskalender}
  \Description{Participation-weighted Mean Attempts (left) and Pass Rates (right) across seven AI competence clusters. The comparison reveals a shift from low-success "bottlenecks" in 2024 (e.g., Temporal) to high-engagement success in 2025, with the notable exception of Reliability.}
  \label{fig:attemptsAndPassRateAcrossClusters}
\end{figure}

The longitudinal data from the 24-day advent period reveals distinct behavioral shifts between the two cohorts, primarily driven by the transition from a blocked to an interleaved curriculum. In the 2024 run, the pedagogical structure resulted in specific technical bottlenecks, most notably within the Temporal and System Level categories, where pass rates plummeted to as low as 10.2\% on certain days despite consistent user effort. Conversely, the 2025 iteration demonstrates a more resilient success profile; although initial questions in the Introduction phase required a higher mean attempt rate (1.47 per user) compared to the previous year, this investment yielded significantly higher success, with pass rates nearly doubling to 81.8\%. Figure \ref{fig:meanAttemptsCategory} shows the mean attempts per competence cluster.

In 2024, System Level and Temporal topics emerged as the most significant hurdles, characterized by low pass rates (33.6\% and 10.2\%, respectively) that did not improve with repeated attempts. The 2025 data shows a remarkable recovery in these areas, with System Level pass rates climbing to 64.5\%, suggesting improved alignment of question difficulty. However, the Reliability category serves as a notable outlier in this trend; despite maintaining a steady attempt rate, its pass rate collapsed from a high of 71.3\% in 2024 to 16.8\% in 2025. Figure \ref{fig:passRateCategory} shows the pass rate based on the competence cluster.

\begin{figure}
  \includegraphics[width=0.65\textwidth]{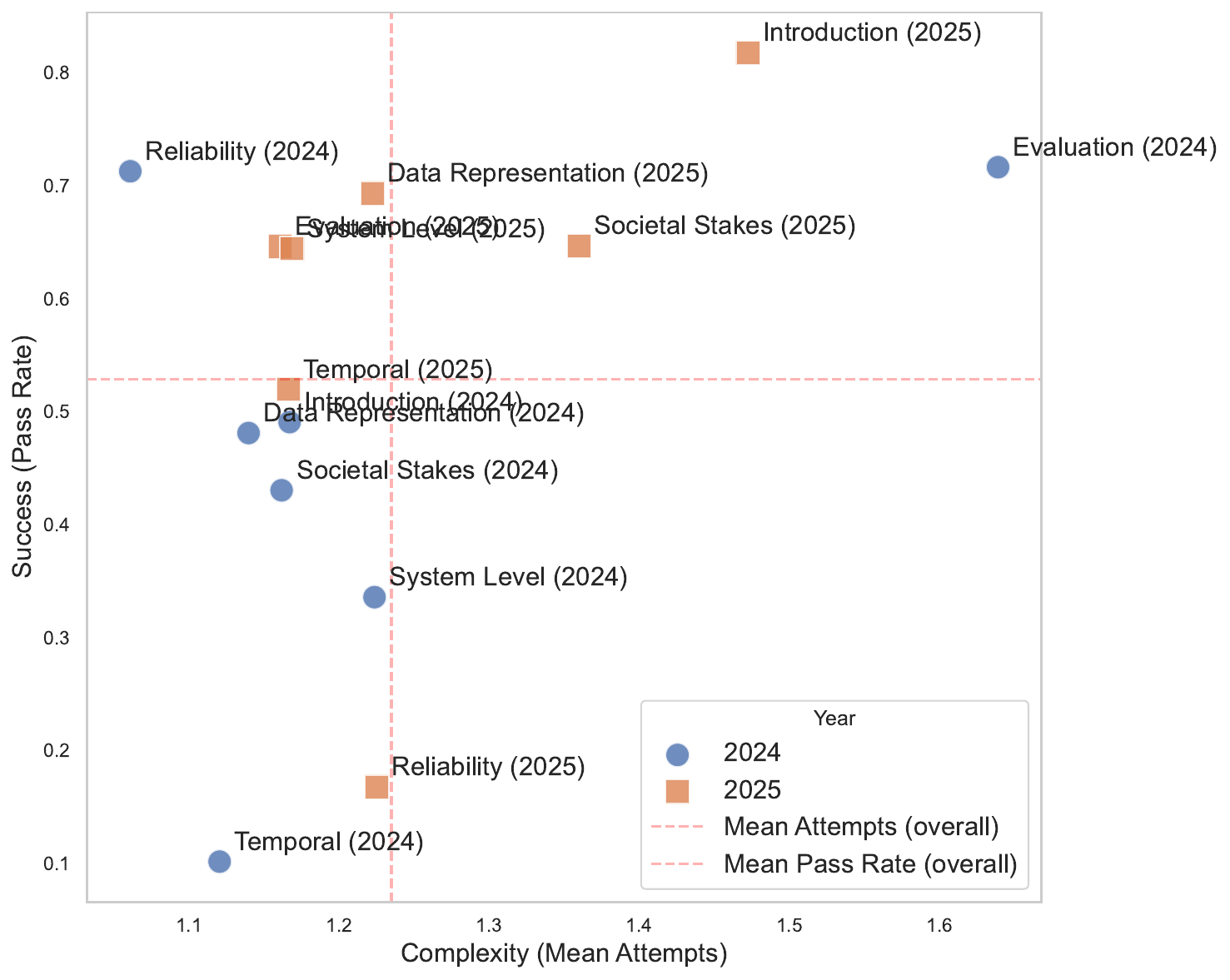}
  \caption{Complexity vs. Success Matrix: Comparison of weighted mean attempts and pass rates by topic category for the 2024 and 2025 curriculum designs}
  \Description{Categorical mapping of user engagement: A complexity-success matrix comparing the 2024 and 2025 KI-Adventskalender runs. Complexity is represented by the participation-weighted mean attempts per question, while success is measured by the aggregate pass rate. The red dashed lines indicate the grand means for both axes across the study period.}
  \label{fig:matrix}
\end{figure}

By utilizing mean attempts as a proxy for complexity and pass rates as a measure of successful engagement, the user experience can be categorizes into four distinct quadrants as shown in Figure \ref{fig:matrix}. In 2024, several key topics were situated in the "High Complexity / Low Success" quadrant, a position typically associated with user frustration and cognitive overload. The 2025 run successfully shifted the majority of these categories toward the "High Success" zone, creating a more rewarding effort-to-outcome ratio. Specifically, the Introduction and Societal Stakes modules in 2025 represent the optimal "Engaging" quadrant, where increased user effort (higher attempts) was consistently rewarded with high pass rates.

\section{Discussion}

\paragraph{From Behavioral Traces to Data Literacy Constructs}
Platform logs capture participation, progression and task performance but these signals may conflate multiple factors. High progression may indicate interest, low barriers or self-selected motivated users. Pass rates reflect task difficulty, prior knowledge, wording and interface design---not conceptual understanding. Revision counts shows \textit{where} learners engage but not \textit{how} they reason about data quality, bias or metric trade-offs. The field needs assessments that distinguish task performance from transferable durable learning \cite{Brockbank2025,10295997}, especially in informal settings.

\paragraph{Design Principles for Data/AI Learning Interventions} 

Two structural choices shaped engagement. First, curriculum sequencing: 2024 blocked doors by topic (e.g., all System Level consecutive), creating bottlenecks---Temporal and System Level showed 10.2\% and 33.6\% pass rates despite repeated attempts. 2025 interleaved topics daily, forcing learners to shift frames (fairness$\rightarrow$embeddings$\rightarrow$drift). This interleaved structure introduced a higher initial cognitive load, resulting in a steeper Day 1$\rightarrow$2 filtering effect,  but systematically raised cluster pass rates for persisters (Figure~\ref{fig:passRateCategory}). Second, interactive elements (threshold sliders, draggable clustering, deepfake comparisons) appeared in doors with higher attempts and higher pass rates in 2025 (Introduction, Societal Stakes, Figure~\ref{fig:matrix}), suggesting interactivity supported productive effort. These features matter for scaling: micro-challenges that sustain engagement without instructor support become viable classroom complements.


\paragraph{Three Forms of Data-Centric Reasoning across Competence Clusters}
The door set operationalizes data literacy as three modes. \textbf{Representational} reasoning (Data Representation cluster) translates between encodings: pixels as binary arrays, text as embeddings, \textit{n}-grams as frequency tables, surfacing the abstraction gap between raw data and model inputs.\textbf{Evaluative} reasoning (Evaluation cluster) interprets metrics and trade-offs: confusion matrices, precision-recall curves, validation splits. Performance claims depend on metric choice and decision context \cite{huey2023communicative}, not algorithms alone. \textbf{Normative} reasoning (Societal Stakes cluster) frames fairness as contestable definitions, privacy as data governance, synthetic media as epistemic risk. Data literacy is critical evaluation of power and bias, not just technical skill \cite{10.1145/3392063.3394396,10.1145/3613904.3642460}. The Reliability cluster collapse in 2025 (71.3\% → 16.8\% pass rate, Figure~\ref{fig:passRateCategory}) raises a question: did learners struggle with XAI method comparison because explanations are inherently difficult or because the task required integrating representational and evaluative reasoning simultaneously? Without finer instrumentation, embedded checks, delayed transfer tasks we cannot currently tell.



\paragraph{Measurement Validity and Cross-Disciplinary Assessment Challenges.}




Our metrics are observational proxies, not validated literacy measures. This echoes challenges across visualization \cite{6875906,10.1145/3544548.3581406}, statistical \cite{gould2017data} and AI literacy \cite{lintner2024systematic} assessments: no consensus on what constitutes evidence of understanding. HCI researchers ask what interaction patterns signal comprehension versus surface engagement. Cognitive scientists ask what task features elicit transferable schemas vs rote recall. Learning scientists ask how to measure durable change in uncontrolled settings. Our contribution is not resolving these questions but illustrating their \textit{interdependence}. The workshop's call for "shared vocabulary" and "bridging related literacies" is urgent: without it, each discipline optimizes local metrics (chart-reading speed, confusion matrix recall, fairness selection) while \textit{integrative reasoning} remains unmeasured.

\subsection{Limitations}
This analysis uses an observational design with self-selection participants. Those who register and persist likely differ from the broader student population in motivation, prior exposure or digital access. Beyond conceptual understanding, task correctness reflects multiple factors: prior knowledge, language proficiency, wording clarity and interface design. Progression may conflate engagement with calendar format novelty, peer effects or teacher encouragement (unobserved in our data).Platform logs provide no evidence of \textit{how} learners reasoned during attempts: whether they guessed, consulted external resources, discussed with peers or reflected on feedback. These are known limitations of trace-based learning analytics \cite{rakovic2023learning}.

\subsection{Future Directions}


To move beyond observational description, future iterations should integrate tighter instrumentation and validation methodologies. \textbf{Embedded micro-assessments}: short comprehension checks after each door (e.g., "\textit{Explain why threshold choice affects precision}") would separate task completion from conceptual grasp. \textbf{Process tracing}: logging interaction sequences (e.g. time on diagrams, slider adjustments prior to submission) could reveal reasoning strategies. Delayed transfer tasks in which participants are asked to apply previously learned logic or concept to a new domain one week later would test progression beyond immediate performance. \textbf{Mixed-methods probes} such as exit surveys asking "\textit{What was hardest about this door?}" would provide context for quantitative patterns. \textbf{Controlled comparisons} such as randomizing users to blocked vs interleaved sequences within a cohort would isolate curriculum effects from year-specific confounds. These additions would support stronger claims about what informal micro-challenges can and cannot teach about data literacy, addressing the workshop's call for "tools that uncover cognitive processes" and "assessments in different contexts."

\section{Conclusion}

\textit{\textbf{KI-Adventskalender}} demonstrates that informal micro-challenge learning interventions can sustain engagement and reveal task-level difficulty patterns across data and AI competence clusters. Across two annual deployments, we observed early curriculum filtering effects followedby robust mid-point persistence, distinct difficulty signatures across competence clusters, and marked performance shifts following a transition from blocked to interleaved sequencing. In particular, improvements in System Level and Temporal clusters contrasted with a sharp decline in Reliability, underscoring that different forms of data reasoning respond differently to structural design choices. However, platform traces alone are insufficient for validating whether participants develop durable data literacy, defined as the ability to reason about representations, evaluation trade-offs and normative stakes in integrated ways. Moving forward, future research should prioritize assessment designs that bridge behavioral proxies and conceptual understanding, shared vocabularies for cross-disciplinary measurement and instrumentation that reveals \textit{how} learners think with data, not just whether they complete tasks.

\section{Acknowledgments}
KI-Adventskalender is a non-profit initiative by German Research Center for Artificial Intelligence and RPTU Kaiserslautern-Landau. We thank our sponsors for supporting the program through prizes and logistical assistance. Sponsors had no influence on the educational content, the running of KI-Adventskalender, data collection, analysis, or interpretation of findings, and were granted no access to participant-level or aggregated data.

\bibliographystyle{ACM-Reference-Format}
\bibliography{sample-base}






\end{document}